\documentclass[a4paper,12pt]{article}

\usepackage{ifpdf}

\newif\ifpdf
\ifx\pdfoutput\undefined
  \pdffalse
\else
  \pdfoutput=1
  \pdftrue
\fi

\RequirePackage{xspace} %
\RequirePackage{subfigure} %
\RequirePackage[centertags]{amsmath} %
\RequirePackage{amssymb}
\RequirePackage{wrapfig} %
\RequirePackage{calc} %
\RequirePackage{ifthen}
\RequirePackage{tabularx} %
\RequirePackage{flafter} %
\RequirePackage{fancyhdr} %

\ifpdf
  \RequirePackage[pdftex]{color}%
  \RequirePackage{colortbl}%
  \RequirePackage{array}%
  \RequirePackage[pdftex]{graphicx}

  \RequirePackage[ pdftex, plainpages = false, pdfpagelabels,
                 pdfpagelayout = useoutlines,
                 bookmarks,
                 breaklinks = true,
                 linktocpage,
                 pagebackref,                      
                 colorlinks = true,
                 linkcolor = blue,
                 urlcolor  = blue,
                 citecolor = blue,
                 anchorcolor = blue,
                 hyperindex = true,
                 hyperfigures
                 ]{hyperref}

\else
  \RequirePackage{color}
  \RequirePackage{colortbl}
   \RequirePackage{array}
  \RequirePackage[dvips]{graphicx}
  \RequirePackage{hyperref}
  \usepackage{rotating}
\fi

\usepackage{makeidx} 
\usepackage{setspace} 
\usepackage{rotating} 
\usepackage{ecltree}
\usepackage{epic}
\usepackage{supertabular}  
\usepackage{color}
\usepackage{exscale}
\usepackage{fontenc}
\usepackage{ifthen}
\usepackage{latexsym}
\usepackage{makeidx}
\usepackage{syntonly}
\usepackage{inputenc}
\usepackage{graphicx}
\usepackage{setspace}
\usepackage{caption2}
\usepackage[english]{babel}
\usepackage[square, comma,numbers,sort&compress]{natbib}
\usepackage{hypernat}
\usepackage{boxedminipage}
\usepackage{framed}
\usepackage{longtable}
\usepackage[all]{hypcap}    
\usepackage{algorithm2e}
\usepackage{algorithmic}
\usepackage{lscape}
\usepackage{pdflscape}

\newcommand{\note}[1]{\marginpar[left]{\singlespace \tiny #1}}
\newcommand{\pois}{Poiseuille}
\newcommand{\Hs}      {\hspace{-0.5cm}} %
\newcommand{\CIF}     {\centering \includegraphics[width=2.7in]} %
\newcommand{\etal}    {{\em et al}} %

\renewcommand{\sectionmark}[1]%
      {\markright{\thesection\ #1}} 

\renewcommand{\note}[1]{}

\onehalfspace 

\begin{document}
\begin{center}
{\Large The Flow of Newtonian and power law fluids in elastic tubes}
\par\end{center}{\Large \par}

\begin{center}
Taha Sochi
\par\end{center}

\begin{center}
{\scriptsize University College London, Department of Physics \& Astronomy, Gower Street, London,
WC1E 6BT \\ Email: t.sochi@ucl.ac.uk.}
\par\end{center}

\begin{abstract}
\noindent We derive analytical expressions for the flow of Newtonian and power law fluids in
elastic circularly-symmetric tubes based on a lubrication approximation where the flow velocity
profile at each cross section is assumed to have its axially-dependent characteristic shape for the
given rheology and cross sectional size. Two pressure-area constitutive elastic relations for the
tube elastic response are used in these derivations. We demonstrate the validity of the derived
equations by observing qualitatively correct trends in general and quantitatively valid asymptotic
convergence to limiting cases. The Newtonian formulae are compared to similar formulae derived
previously from a one-dimensional version of the Navier-Stokes equations.

\vspace{0.3cm}


\par\end{abstract}

\section{Introduction} \label{Introduction}

The flow of Newtonian and non-Newtonian fluids in distensible conduits of elastic or viscoelastic
nature is common in many biological and industrial systems. Some examples are the flow of
biological fluids such as blood in the vessels and porous tissues of most organisms including
human, the transportation of drainage extracts, and the delivery of various liquid products in the
oil and food industries.

A number of attempts have been made in the past to model the flow of fluids in distensible
conduits; most of which are restricted to Newtonian fluids and relatively few considered
non-Newtonian rheologies. An example of the former is the widely used one-dimensional (1D)
Navier-Stokes flow model in deformable tubes \cite{BarnardHTV1966, BarnardHTV1966b, Olufsen2001,
SmithPH2002, FormaggiaLQ2003, SherwinFPP2003}, while an example of the latter is the attempt of
Vajravelu \etal\ \cite{VajraveluSDP2011} to model the flow of Herschel-Bulkley fluids in elastic
tubes. Other attempts have also been made to deal with more special cases of non-Newtonian flow
using mainly numerical methods \cite{Mandal2005, SreedharamallePRK2012}.

Although the 1D Navier-Stokes distensible flow model is widely used in the flow simulations in
single distensible tubes and networks of interconnected distensible tubes especially in the
hemodynamic studies, it has a number of limitations. One of these limitations is its restriction to
the Newtonian flow, on which the Navier-Stokes momentum equation is based, and hence to accommodate
non-Newtonian rheologies, the use of approximations or employing different models, if they exist,
is required. Another limitation is the relatively large number of parameters (e.g. $\alpha$,
$\kappa$, $\rho$, etc.) that define the 1D model; which make it difficult to choose proper
numerical values for these parameters in practical situations without considerable amount of
experimental and observational work, and hence some ambiguity or arbitrariness may be attached to
the assumptions and results of this model. This has been demonstrated in \cite{SochiPois1DComp2013}
where the results of the 1D Navier-Stokes flow model can vary considerably on varying these
parameters; and hence the model may loose its predictive value due to large uncertainties in the
results as a consequence of uncertainties in the input parameters.

Concerning the non-Newtonian case, the approaches developed to deal with various non-Newtonian
rheologies in distensible conduits face serious analytical and numerical difficulties; hence
hindering their development and usage and limiting their formulation and application to rather
special cases with considerable approximations. A quick review to the existing literature will
reveal that limited progress has been made on most aspects of the non-Newtonian flow in deformable
conduits. There are many gaps in the research literature on this subject where no attempts have
been made to investigate some commonplace problems.

Based on this assessment to the existing Newtonian and non-Newtonian flow models, it can be
concluded that it is highly desirable to have flow models with comparatively few parameters that
can describe both Newtonian and non-Newtonian rheologies even if they are developed by the
employment of simpler approaches. Although such models may seem idealized, they can be used in many
practical situations; moreover, they can serve as prototypes for the development of more
sophisticated models.

In the present paper, we try to develop one such model for the Newtonian and power law fluids using
two pressure-area elastic relations to describe the tube distensibility. We use a lubrication
approximation approach based on assuming a fluid-specific velocity profile that varies with the
size of the cross sectional area which can be assumed to vary smoothly in the axial direction for
the investigated case of laminar flow through distensible conduits. We partially validate the
derived analytical expressions by observing a number of sensible trends mostly related to
convergence to limiting cases. We also propose the use of these expressions in conjunction with the
previously proposed \cite{SochiPoreScaleElastic2013} residual-based pore-scale network modeling to
simulate the flow of Newtonian and power law fluids in networks of interconnected distensible tubes
that can also be used as models for porous media.

Extending the method to rheologies other than Newtonian and power law fluids and to conduits of
distensible nature other than the two simple pressure-area elastic models is a possibility
especially if numerical, as well as analytical, methods are considered. Also, broadening the
proposed method to include conduit geometries other than the regular cylindrical tube with a
constant cross sectional area in the axial direction (e.g. conduits of elliptically-shaped cross
sections or of converging-diverging nature) may also be possible; although we will not consider any
of these extensions in the present paper.

\section{Pressure-Area Relations}

There are many constitutive models that correlate the local pressure to the local cross sectional
area in distensible tubes; these models include elastic and viscoelastic mechanical response, and
linear and nonlinear correlation, as well as many other more specific variations. In this paper we
use two elastic models to demonstrate the use of the proposed method.

One of these is a simple model in which a linear pressure-area correlation is assumed, that is

\begin{equation}\label{pA1}
p=\gamma\left(A-A_{o}\right)
\end{equation}
where $\gamma$ is the proportionality factor which is a measure for the stiffness of the tube wall,
$A$ is the tube cross sectional area at the actual pressure $p$ as opposite to the reference
pressure which, for the sake of convenience without loss of generality, is assumed to be zero, and
$A_{o}$ is the reference area corresponding to the reference pressure. The core of this model is
that the locally defined axial pressure is proportional to the change in the local cross sectional
area relative to its reference state, where, for all the cases considered in this paper, we assume
$A\ge A_{o}$ to exclude the collapsible tube case.

The other pressure-area model is based on the proportionality between the local pressure and the
change in the local radius relative to its reference state with a proportionality stiffness factor
that is scaled by the reference area, i.e.

\begin{equation}\label{pA2}
p=\frac{\beta}{A_{o}}\left(\sqrt{A}-\sqrt{A_{o}}\right)
\end{equation}
where $\beta$ is the tube stiffness factor.

In the next section we use these two pressure-area models to derive flow relations for Newtonian
and power law fluids using the proposed lubrication method.

\section{Derivation}

In this section we derive analytical relations for the volumetric flow rate, $Q$, as a function of
the pressure boundary conditions in distensible tubes for the Newtonian and power law rheologies
using the two pressure-area elastic constitutive relations given in the preceding section. In these
derivations we assume an incompressible, time-independent, laminar, fully developed flow at
relatively low Reynolds numbers with minimal entry and exit edge effects. The impact of edge
effects should be reduced for sufficiently long tubes in laminar flow conditions. We also assume
that the elastic tube has a fixed circularly-shaped cross section with a constant cross sectional
area in the axial direction under unstressed state conditions. Another assumption is that the tube
has a constant length and hence any stretch in the axial direction under stressed state conditions
is negligible. The vessel wall is normally assumed to be thin, homogeneous, isotropic, of constant
thickness with linear distensibility and negligible compressibility \cite{FormaggiaNQV1999,
AlastrueyMPDPS2007, JanelaMS2010}; and we follow suit. The essence of these conditions is that the
same pressure-area relation applies at all cross sections producing an axi-symmetric change in the
tube geometry.

\subsection{Newtonian Fluids}

For Newtonian fluids the rheology is given by the following stress-strain relation

\begin{equation}
\tau=\mu\dot{\gamma}
\end{equation}
where $\tau$ is the shear stress, $\mu$ is the fluid dynamic viscosity, and $\dot{\gamma}$ is the
rate of shear strain. Based on the Hagen-\pois\ flow assumptions, the volumetric flow rate, $Q$, is
correlated to the fluid dynamic viscosity, $\mu$, tube radius, $r$, and pressure gradient
$\frac{dp}{dx}$ through the following relation

\begin{equation}\label{poisQ}
Q=\frac{\pi r^{4}}{8\mu}\frac{dp}{dx}=\frac{A^{2}}{8\pi\mu}\frac{dp}{dx}
\end{equation}
where $x$ is the tube axial coordinate, $p$ is the pressure in the axial direction and $A$ is the
cross sectional area. If we assume, as a consequence of the validity of the aforementioned
assumption of flow profile dependency, that such a relation is valid at each cross section even for
an elastic tube with variable cross sectional area that is subject to an axially-dependent pressure
as long as the variation in the pressure, and hence the corresponding cross sectional area, is
smooth, we then have the following relation at each cross section

\begin{equation}
\frac{dp}{dx}=\frac{8\pi\mu Q}{A^{2}}\label{dpdxNewt1}
\end{equation}
where $Q$ as a function of the axial coordinate, $x$, is constant due to the incompressibility and
time-independent flow conditions.

Now, from the first $p$-$A$ relation we have

\begin{equation}
A=\frac{p}{\gamma}+A_{o}
\end{equation}
and hence from Equation \ref{dpdxNewt1} we obtain

\begin{equation}
\frac{dp}{dx}=\frac{8\pi\mu Q}{\left(\frac{p}{\gamma}+A_{o}\right)^{2}}\label{dpdxNewt2}
\end{equation}
which can be separated and integrated, i.e.

\begin{equation}
\int\left(\frac{p}{\gamma}+A_{o}\right)^{2}dp=\int8\pi\mu Qdx
\end{equation}
to yield the following relation

\begin{equation}
\frac{\left(p+\gamma A_{o}\right)^{3}}{3\gamma^{2}}=8\pi\mu Qx+C
\end{equation}
where $C$ is the constant of integration. From the inlet boundary condition, we have
$p\left(x=0\right)=p_{in}$ where $p_{in}$ is the pressure at the tube inlet, and hence

\begin{equation}
C=\frac{\left(p_{in}+\gamma A_{o}\right)^{3}}{3\gamma^{2}}
\end{equation}
that is

\begin{equation}
Q=-\frac{\left(p_{in}+\gamma A_{o}\right)^{3}-\left(p+\gamma A_{o}\right)^{3}}{24\pi\mu\gamma^{2}x}
\end{equation}
where the minus sign indicates the fact that the flow is opposite in direction to the pressure
gradient, i.e. $p_{in}>p_{ou}$ where $p_{ou}$ is the pressure at the tube outlet. Applying the
second boundary condition at the outlet, i.e. $p\left(x=L\right)=p_{ou}$ where $L$ is the tube
length, and dropping the minus sign as we are interested only in the magnitude of the flow rate
since its direction is well known (i.e. from inlet to outlet), we obtain

\begin{equation}\label{NewtQ1}
Q=\frac{\left(p_{in}+\gamma A_{o}\right)^{3}-\left(p_{ou}+\gamma
A_{o}\right)^{3}}{24\pi\mu\gamma^{2}L}
\end{equation}

Following a similar procedure and using the second pressure-area relation, i.e. Equation \ref{pA2},
we obtain

\begin{equation}\label{NewtQ2}
Q=\frac{\beta}{40\pi\mu
A_{o}L}\left[\left(\frac{A_{o}}{\beta}p_{in}+\sqrt{A_{o}}\right)^{5}-\left(\frac{A_{o}}{\beta}p_{ou}+\sqrt{A_{o}}\right)^{5}\right]
\end{equation}

\subsection{Power Law Fluids}

For power law fluids the stress-strain rheological relation is given by

\begin{equation}\label{pl}
\tau=k\dot{\gamma}^{n}
\end{equation}
where $k$ is the power law consistency factor and $n$ is the flow behavior index. Based on the
previously given assumptions, the following relation for the volumetric flow rate as a function of
the fluid rheology, tube radius and pressure gradient can be derived
\cite{SochiThesis2007,SochiVariational2013}

\begin{equation}\label{plQ}
Q=\frac{\pi n}{3n+1}\sqrt[n]{\frac{1}{2k}\frac{dp}{dx}}r^{3+1/n}=\frac{\pi
n}{3n+1}\sqrt[n]{\frac{1}{2k}\frac{dp}{dx}}\left(\sqrt{\frac{A}{\pi}}\right)^{3+1/n}
\end{equation}
which can be manipulated to give

\begin{equation}
\frac{dp}{dx}=\frac{2k\pi^{(n+1)/2}\left(3n+1\right)^{n}Q^{n}}{n^{n}A^{(3n+1)/2}}
\end{equation}

On using the first $p$-$A$ relation to replace $A$ with $p$, separating the $p$ and $x$ variables,
integrating and applying the two boundary conditions, as demonstrated in the Newtonian case, we
obtain

\begin{equation}\label{plQ1}
Q=\left(\frac{\gamma
n^{n}\left[\left(\frac{p_{in}}{\gamma}+A_{o}\right)^{3(n+1)/2}-\left(\frac{p_{ou}}{\gamma}+A_{o}\right)^{3(n+1)/2}\right]}{3k\pi^{(n+1)/2}\left(3n+1\right)^{n}\left(n+1\right)L}\right)^{1/n}
\end{equation}

Repeating the process with the use of the second $p$-$A$ relation we obtain

\begin{equation}\label{plQ2}
Q=\left(\frac{\beta
n^{n}\left[\left(\frac{A_{o}}{\beta}p_{in}+\sqrt{A_{o}}\right)^{(3n+2)}-\left(\frac{A_{o}}{\beta}p_{ou}+\sqrt{A_{o}}\right)^{(3n+2)}\right]}{2k\pi^{(n+1)/2}\left(3n+1\right)^{n}\left(3n+2\right)A_{o}L}\right)^{1/n}
\end{equation}

\section{Validation}

\begin{figure} [!h]
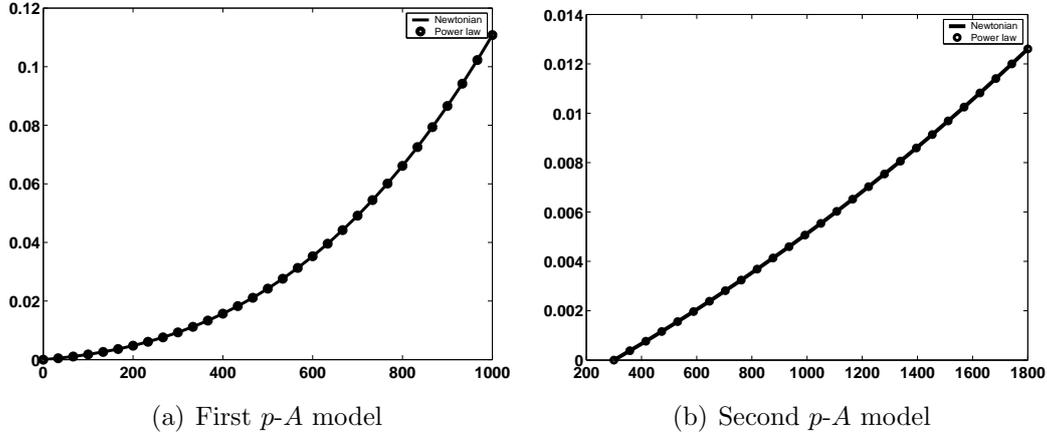

\centering %
\subfigure[First $p$-$A$ model]%
{\begin{minipage}[b]{0.5\textwidth} \CIF {g/PlotConvergencePlToNewt1}
\end{minipage}}
\Hs %
\subfigure[Second $p$-$A$ model]%
{\begin{minipage}[b]{0.5\textwidth} \CIF {g/PlotConvergencePlToNewt2}
\end{minipage}}
\caption{Convergence of the power law formulae (Equations \ref{plQ1} and \ref{plQ2}) when $n=1.0$
to their corresponding Newtonian formulae (Equations \ref{NewtQ1} and \ref{NewtQ2}) for elastic
tubes
(a) using the first $p$-$A$ elastic model (Equation \ref{pA1}) with $\gamma=10^5$~Pa.m$^{-2}$,
$L=0.25$~m, $R_o=0.03$~m, $\mu=k=0.1$~Pa.s, and $p_{ou}=0.0$~Pa; and
(b) using the second $p$-$A$ elastic model (Equation \ref{pA2}) with $\beta=500$~Pa.m, $L=0.35$~m,
$R_o=0.025$~m, $\mu=k=0.075$~Pa.s, and $p_{ou}=300$~Pa.
The vertical axis in the sub-figures represents the volumetric flow rate, $Q$, in m$^3$.s$^{-1}$
while the horizontal axis represents the inlet boundary pressure, $p_{in}$, in Pa.
\label{PlotConvergencePlToNewt}}
\end{figure}

We do not have an independent way, such as experimental data, to validate the derived formulae.
However, we observe a number of sensible trends that can be regarded as partial verification:

\begin{itemize}

\item
For both cases of Newtonian and power law fluids, we observe the convergence of the flow rate as
obtained from the derived formulae to their rigid equivalents with increasing the tube stiffness,
that is the convergence of Equations \ref{NewtQ1} and \ref{NewtQ2} to the \pois\ flow for the
Newtonian fluids, as given by Equation \ref{poisQ}, and the convergence of Equations \ref{plQ1} and
\ref{plQ2} to the power law flow, as given by Equation \ref{plQ}, for a tube with constant radius
and hence constant pressure gradient in the axial direction. In Figures
\ref{PlotConvergenceToRigidPA1} and \ref{PlotConvergenceToRigidPA2} we demonstrate this trend by a
few examples from the Newtonian and power law fluids using the two pressure-area constitutive
elastic relations.

\item
Another validation test is the convergence of the power law formulae for elastic tubes (Equations
\ref{plQ1} and \ref{plQ2}) to their corresponding Newtonian formulae (Equations \ref{NewtQ1} and
\ref{NewtQ2}) when $n=1.0$. In Figure \ref{PlotConvergencePlToNewt} we demonstrate this using two
examples: one from the first $p$-$A$ elastic tube model (Equation \ref{pA1}), and the second from
the second $p$-$A$ elastic tube model (Equation \ref{pA2}). As seen, the Newtonian and power law
formulae, which are derived independently using two different flow relations, produce identical
results in this case as it should be. A consequence of this convergence is that in practical
situations the Newtonian formulae are redundant as they can be obtained as a special case from the
power law formulae.

\item
As a consequence of the previous two points, the power law formulae (Equations \ref{plQ1} and
\ref{plQ2}) converge to the rigid \pois\ flow (Equation \ref{poisQ}) when $n=1.0$ with high tube
stiffness. This trend was verified in all the cases that were investigated.

\item
There are several other qualitatively sensible trends that were verified in the newly derived
formulae. These trends include shear thinning and shear thickening behaviors in the case of power
law fluids, radius and pressure variation in the axial direction, and the magnitude of the flow and
its connection to the average cross sectional area.

\end{itemize}


\begin{figure} [!h]
\centering %
\subfigure[Newtonian]%
{\begin{minipage}[b]{0.5\textwidth} \CIF {g/PlotConvergenceToRigidPA1Newt}
\end{minipage}}
\Hs %
\subfigure[Power law]%
{\begin{minipage}[b]{0.5\textwidth} \CIF {g/PlotConvergenceToRigidPA1PL}
\end{minipage}}
\caption{Convergence of the elastic tube formulae (Equations \ref{NewtQ1} and \ref{plQ1}) to their
rigid equivalents (Equations \ref{poisQ} and \ref{plQ}) at high tube wall stiffness for the first
$p$-$A$ elastic tube model (Equation \ref{pA1})
(a) for a Newtonian fluid with $\mu=0.05$~Pa.s; and
(b) for a shear thinning power law fluid with $k=0.05$~Pa.s$^n$, and $n=0.75$.
The tube parameters are: $\gamma=10^8$~Pa.m$^{-2}$, $L=0.5$~m, and $R_o=0.05$~m. The vertical axis
in the sub-figures represents the volumetric flow rate, $Q$, in m$^3$.s$^{-1}$ while the horizontal
axis represents the inlet boundary pressure, $p_{in}$, in Pa. The outlet pressure is held constant
at $p_{ou}=500$~Pa. \label{PlotConvergenceToRigidPA1}}
\end{figure}


\begin{figure} [!h]
\centering %
\subfigure[Newtonian]%
{\begin{minipage}[b]{0.5\textwidth} \CIF {g/PlotConvergenceToRigidPA2Newt}
\end{minipage}}
\Hs %
\subfigure[Power law]%
{\begin{minipage}[b]{0.5\textwidth} \CIF {g/PlotConvergenceToRigidPA2PL}
\end{minipage}}
\caption{Convergence of the elastic tube formulae (Equations \ref{NewtQ2} and \ref{plQ2}) to their
rigid equivalents (Equations \ref{poisQ} and \ref{plQ}) at high tube wall stiffness for the second
$p$-$A$ elastic tube model (Equation \ref{pA2})
(a) for a Newtonian fluid with $\mu=0.15$~Pa.s; and
(b) for a shear thinning power law fluid with $k=0.15$~Pa.s$^n$, and $n=0.6$.
The tube parameters are: $\beta=50000$~Pa.m, $L=0.1$~m, and $R_o=0.015$~m. The vertical axis in the
sub-figures represents the volumetric flow rate, $Q$, in m$^3$.s$^{-1}$ while the horizontal axis
represents the inlet boundary pressure, $p_{in}$, in Pa. The outlet pressure is held constant at
$p_{ou}=200$~Pa. \label{PlotConvergenceToRigidPA2}}
\end{figure}


\section{Comparison to 1D Navier-Stokes Model}

In \cite{SochiElastic2013} two formulae based on the widely used 1D Navier-Stokes model for the
flow of Newtonian fluids in elastic tubes were derived. These formulae, which are based on the
first and second $p$-$A$ relations respectively, are given by

\begin{equation}\label{QElastic1}
Q=\frac{L-\sqrt{L^{2}-4\frac{\alpha}{\kappa}\ln\left(A_{ou}/A_{in}\right)\frac{\gamma}{3\kappa\rho}\left(A_{in}^{3}-A_{ou}^{3}\right)}}{2\frac{\alpha}{\kappa}\ln\left(A_{ou}/A_{in}\right)}
\end{equation}
and

\begin{equation}\label{QElastic2}
Q=\frac{-\kappa L+\sqrt{\kappa^{2}L^{2}-4\alpha\ln\left(A_{in}/A_{ou}\right)\frac{\beta}{5\rho
A_{o}}\left(A_{ou}^{5/2}-A_{in}^{5/2}\right)}}{2\alpha\ln\left(A_{in}/A_{ou}\right)}
\end{equation}
where $\alpha$ is the axial momentum flux correction factor which is related to the flow velocity
profile, $\kappa$ is a viscosity friction coefficient, $\rho$ is the fluid mass density, and
$A_{in}$ and $A_{ou}$ are the inlet and outlet cross sectional area of the tube respectively.

\begin{figure} [!h]
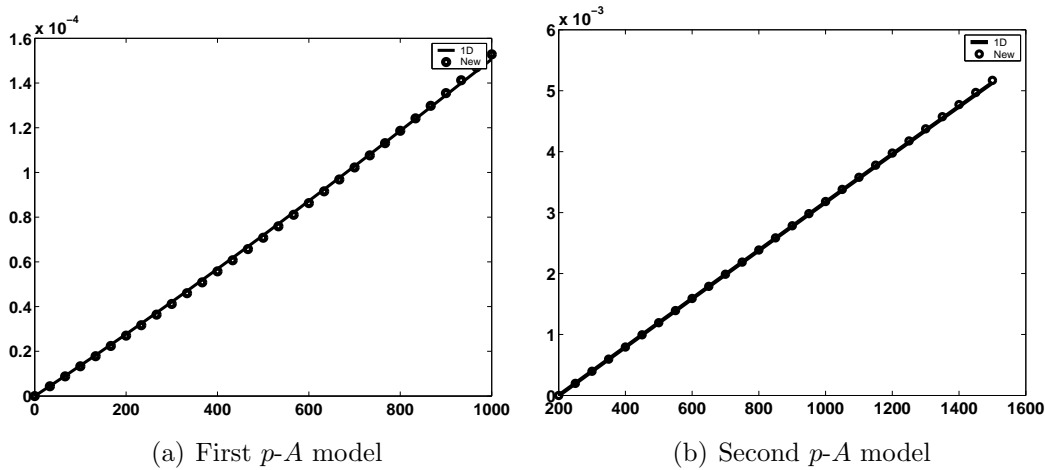

\centering %
\subfigure[First $p$-$A$ model]%
{\begin{minipage}[b]{0.5\textwidth} \CIF {g/Plot1DTunePA1}
\end{minipage}}
\Hs %
\subfigure[Second $p$-$A$ model]%
{\begin{minipage}[b]{0.5\textwidth} \CIF {g/Plot1DTunePA2}
\end{minipage}}
\caption{Comparing the 1D Navier-Stokes model formulae (Equations \ref{QElastic1} and
\ref{QElastic2}) with the newly derived formulae (Equations \ref{NewtQ1} and \ref{NewtQ2}) for the
flow of Newtonian fluids
(a) using the first $p$-$A$ elastic model (Equation \ref{pA1}) with $\alpha=1.345$,
$\gamma=2.0\times 10^7$~Pa.m$^{-2}$, $\mu=0.3$~Pa.s, $\rho=1000.0$~kg.m$^{-3}$, $L=0.1$~m,
$R_o=0.01$~m, and $p_{ou}=0.0$~Pa; and
(b) using the second $p$-$A$ elastic model (Equation \ref{pA2}) with $\alpha=1.333$,
$\beta=5.0\times 10^5$~Pa.m, $\mu=0.05$~Pa.s, $\rho=900.0$~kg.m$^{-3}$, $L=0.1$~m, $R_o=0.015$~m,
and $p_{ou}=200$~Pa.
The vertical axis in the sub-figures represents the volumetric flow rate, $Q$, in m$^3$.s$^{-1}$
while the horizontal axis represents the inlet boundary pressure, $p_{in}$, in Pa.
\label{Plot1DTunePA}}
\end{figure}

It is difficult to compare the newly derived formulae with the old ones which are obtained from the
1D Navier-Stokes model. First no comparison can be made for the power law fluids because the 1D
Navier-Stokes model is restricted to Newtonian fluids. As for the Newtonian case, the two models
are based on different formulation frameworks and hence it is difficult to make a fair comparison
between the two. One major difficulty is that the 1D Navier-Stokes model contains several
parameters ($\alpha$, $\kappa$ and $\rho$) which are absent in the newly derived formulae. In
particular, $\alpha$ is strongly dependent on the flow velocity profile and hence can cause major
discrepancy with the new formulae which are based essentially on a specific fluid-dependent
velocity profile. In \cite{SochiPois1DComp2013} it has been demonstrated that the 1D model can be
very sensitive to the variation in its parameters. However, in certain cases the 1D model can have
a very close match with the newly derived formulae by adjusting the 1D parameters. For
demonstration purposes, we produced two examples, seen in Figure \ref{Plot1DTunePA}, where we
compared the 1D formulae with the new formulae for the Newtonian flow using the two $p$-$A$
relations. In these examples, we `tuned' the 1D parameters to match the flow as predicted by the
new formulae. The 1D model can also be tuned to match the prediction of the new formulae using sets
of parameters other than those used in these examples.


One of the advantages of the newly derived formulae over the formulae of the 1D Navier-Stokes model
is that the new formulae are simpler than the 1D ones as they contain less parameters. The many
parameters of the 1D model, although may be a source of more diversity in the flow modeling, can
reduce the predictive capability of the model due to the difficulty of obtaining accurate values
for the model parameters and hence the model merit may be relegated to a descriptive state by
`tuning' the parameters to match the observed flow.

Finally, it should be remarked that the newly derived formulae can be used in modeling and
simulating the flow in networks of interconnected elastic tubes by using the pore-scale network
modeling approach as proposed in \cite{SochiPoreScaleElastic2013}. An obvious advantage of the use
of the newly derived formulae over the 1D formulae in a pore-scale modeling scheme is that the new
formulae can describe the flow of non-Newtonian fluids of power law type which the 1D Navier-Stokes
model cannot do because of its restriction to the Newtonian fluids. Other types of non-Newtonian
fluids with other types of distensibility models, such as viscoelastic or elastic with different
$p$-$A$ constitutive relations to those used in the present paper, can also be used in pore-scale
modeling if pertinent analytical or empirical formulae were obtained.

\section{Conclusions} \label{Conclusions}

In this paper we derived analytical formulae for the volumetric flow rate as a function of the two
pressure boundary conditions, the fluid rheology, and the duct geometry for the flow of Newtonian
and power law fluids in distensible tubes of regular cylindrical shapes. In these derivations we
used two pressure-area constitutive relations of elastic nature. The method is based on a
lubrication approximation where the flow velocity profile is assumed to be determined locally by
the fluid rheology and the size of the local cross sectional area as in the case of the flow in a
tube with a cross sectional area that, under flow state, is constant in shape and size over the
whole tube length.

The Newtonian formulae can be used as an alternative to the previously derived formulae
\cite{SochiElastic2013} which are based on the 1D Navier-Stokes flow in distensible tubes. As well
as their possible use in modeling the flow in single distensible tubes, the derived expressions can
be used in modeling the flow in elastic networks, as an alternative to the widely used 1D
Navier-Stokes finite element method \cite{SochiTechnical1D2013}, in conjunction with the
previously-proposed \cite{SochiPoreScaleElastic2013} pore-scale modeling approach. The newly
derived power law formulae will facilitate modeling non-Newtonian rheology in single distensible
tubes and networks of interconnected distensible tubes by the use, in the latter case, of
pore-scale modeling. This is a major addition to the existing modeling capabilities which are
limited to the Newtonian rheology.

For the purpose of validation, several sensible trends have been observed. These include (a) the
convergence of the derived formulae to their corresponding rigid tube formulae with increasing the
stiffness of the tube wall, and (b) the convergence of the power law formulae to their Newtonian
equivalents when the power law index, $n$, is set to unity. Thorough tests have revealed that the
newly derived formulae produce mathematically and physically sensible results in diverse situations
of fluid rheology, tube geometry and boundary conditions.

A brief comparison has been made with the Newtonian formulae that were derived previously from the
1D Navier-Stokes distensible model. A major advantage of the newly derived formulae over the 1D
Navier-Stokes formulae is the accommodation of the non-Newtonian rheology in the form of the power
law model.

The derivation method proposed in this paper can in principle be extended to less regular
geometries (e.g. converging-diverging) or to regular geometries but with non-cylindrical shape
(e.g. square or elliptic cross section) although empirical or numerical, rather than analytical,
approaches may be needed. Similarly, the method may also be extended to fluid rheologies other than
Newtonian and power law fluids with possible restriction to the use of empirical or numerical,
instead of analytical, approaches due to potential mathematical difficulties.

\vspace{1cm}
\phantomsection \addcontentsline{toc}{section}{Nomenclature} %
{\noindent \LARGE \bf Nomenclature} \vspace{0.5cm}

\begin{supertabular}{ll}
$\alpha$                &   correction factor for axial momentum flux in the 1D model \\
$\beta$                 &   stiffness factor in the second pressure-area model \\
$\gamma$                &   stiffness factor in the first pressure-area model \\
$\dot{\gamma}$          &   shear strain rate \\
$\kappa$                &   viscosity friction coefficient \\
$\mu$                   &   fluid dynamic viscosity \\
$\rho$                  &   fluid mass density \\
$\tau$                  &   shear stress \\
\\
$A$                     &   tube cross sectional area corresponding to pressure $p$ \\
$A_{in}$                &   tube cross sectional area at inlet \\
$A_o$                   &   tube reference cross sectional area corresponding to reference pressure \\
$A_{ou}$                &   tube cross sectional area at outlet \\
$k$                     &   power law consistency factor \\
$L$                     &   tube length \\
$n$                     &   power law flow behavior index \\
$p$                     &   axial pressure \\
$p_{in}$                &   pressure at tube inlet \\
$p_{ou}$                &   pressure at tube outlet \\
$Q$                     &   volumetric flow rate \\
$r$                     &   tube radius \\
$R_o$                   &   tube radius corresponding to $A_o$  \\
$x$                     &   tube axial coordinate \\
\end{supertabular}

\phantomsection \addcontentsline{toc}{section}{References} %
\bibliographystyle{unsrt}

\end{document}